\documentclass[twocolumn,superscriptaddress,amsmath,ammsymb]{revtex4-1}
\usepackage{epsfig}
\usepackage{amsmath}
\usepackage{amssymb}
\usepackage{graphicx}
\usepackage{dcolumn}
\usepackage{bm}
\usepackage[usenames]{color}
\usepackage[breaklinks]{hyperref}
\usepackage{soul}
\usepackage{ulem}
\hypersetup{colorlinks=true,allcolors=blue}

\begin{document}
	
\title{Two-dimensional bound excitons in real space and Landau quantization space: A comparative study}  
\author{Kunxiang Li}
\affiliation{School of Optoelectronic Information and Physical Science, Jiangnan University, Wuxi 214122, China}
\author{Yi-Xiang Wang}
\email{wangyixiang@jiangnan.edu.cn}
\affiliation{School of Optoelectronic Information and Physical Science, Jiangnan University, Wuxi 214122, China}
\date{\today}

\begin{abstract} 
The Landau quantization space is based on the respective motion of the electron and hole in a magnetic field and can provide a new route to understand the bound exciton behaviors observed in the experiments.  
In this paper, we study the two-dimensional exciton properties of monolayer WSe$_2$ in both real space and Landau quantization space.  Focusing on the excitons of zero center-of-mass momentum, we calculate the energy spectrum in both spaces, with the results agreeing well with each other.  We then obtain the diamagnetic coefficients and root-mean-square radius, which are consistent with the available $s$ state data in the experiments.  More importantly, in the exciton state $nl$, we find that the dominant electron-hole pair component may shift with the magnetic field and the Coulomb interactions, and reveal that the magnetic field will drive the dominant component to the free electron-hole pair $\{n_e=n+l-1,n_h=n-1\}$, whereas the Coulomb interactions drives it to the pair of the lower index.     
\end{abstract}  

\maketitle

\section{Introduction} 

Recently, the monolayer transition-metal dichalcogenides (TMDs) have attracted tremendous research interests due to their remarkable properties that show potential applications in optoelectronics and valleytronics~\cite{X.Xu, K.F.Mak}.  In monolayer TMDs, they have direct band gaps in the low-energy valleys of the hexagonal Brillouin zone~\cite{M.Donck2018} and the broken inversion symmetry~\cite{D.A.Ruiz}.  The latter, when combined with the strong spin-orbit couplings, will lead to the lockup of the spin and valley degree of freedom~\cite{D.Xiao}.  
Moreover, the strong light-matter couplings are allowed and will induce the optical excitations of the exciton states that are made up of the electron-hole pair~\cite{G.Wang}. 
In such 2D systems, the reduced dimension will yield a significant enhancement of  the Coulomb interactions between the electrons and holes~\cite{N.S.Rytova, Keldysh, P.Cudazzo}.  Consequently, the excitons will be tightly bound and can have a much larger binding energy than those in conventional two-dimensional (2D) semiconductors, such as the single GaAs quantum well~\cite{M.Hayne} and GaAs/AlGaAs quantum wells~\cite{C.Riva}.   

Experimentally, the excitons in monolayer TMDs were identified through the salient excitonic resonances in the photoluminescence or reflectance spectroscopy conducted under magnetic fields~\cite{A.V.Stier2016a, A.V.Stier2016b, A.V.Stier2018, E.Liu2019, E.Liu2020, M.Molas}.  
To unambiguously resolve these exciton states, the high-quality samples are required, which have been markedly improved due to the fabrication technique developments in recent years.  The identifications of the exciton states in the experiments are of certain significance, \textit{e.g.}, they can provide valuable information about the enhanced Coulomb interactions between the electrons and holes and the interaction-driven novel phase in such 2D systems. 

As the magnetic field can polarize the spin and valley degrees of freedom of resident carriers in monolayer TMDs, it has become a useful tool to explore the exciton properties in the experiments~\cite{E.Liu2019, E.Liu2020, J.Li2020, J.Li2022}.  
Under a magnetic field, a free electron or hole will be deflected by the Lorentz force to move on curved orbits with a set of discrete energies, or called the Landau levels (LLs); while a charge-neutral exciton
is not subjected to the Lorentz force.  In Ref.~\cite{D.V.Tuan2025}, the authors proposed a thought-provoking question: how can a bound exciton state under a magnetic field be expressed through the free electron and hole LL components.  
Although real space was commonly used to study the exciton properties~\cite{A.Spiridonova, M.Donck2018, M.Donck2019, M.N.Brunetti}, it cannot solve the question of the exciton state composition.  
On the other hand, the Landau quantization space provides a new route to understand the exciton behaviors, whose basis is composed of the products of the free electron and hole LL wave functions.  In the Landau quantization space, when the Coulomb potential energy is accurately formulated, the above exciton composition question can be well addressed~\cite{D.V.Tuan2025}.  
However, a full comparison of the exciton energy spectrum under a magnetic field in real space and Landau quantization space is still lacking, which motivates the present work.   
Moreover, in an exciton state, it is interesting to explore whether the exciton state composition could be modulated by the external factors, \textit{e.g.}, the magnetic field and Coulomb interactions?  

In this paper, we will make a comparative study of 2D exciton states in monolayer TMDs under a magnetic field in both real space and Landau quantization space.  We take monolayer WSe$_2$ as an example and focus on the excitons of zero center-of-mass momentum, or called the excitons in the light cone.  The calculation methods in the two spaces will be detailed at length in Secs. III A and III B, respectively.  
The obtained main results are given as follows.    
(i) We systematically compare the exciton energy spectrum obtained in the two spaces and find that the results agree well with each other.  We point out that the magnetic quantum number $l$ can be endowed with the meaning of pairing the electrons and holes as $n_e=n_h+l$~\cite{D.V.Tuan2025}, in which $n_{e(h)}$ denotes the electron (hole) LL index.  
(ii) For the diamagnetic shift $\Delta\varepsilon_{nl}^{\text{dia}}$ in the exciton state $nl$, according to the zero-field wave function calculations as well as the quadratic fittings to the energy, we obtain the diamagnetic coefficient $\sigma_{nl}$ and root-mean-square (rms) radius $r_{nl}$.  The results show good agreement with the available $s$ state data in the experiments~\cite{A.V.Stier2018, E.Liu2019}.  From the fittings, we demonstrate that the $\Delta\varepsilon_{nl}^{\text{dia}}$ shows the quadratic behavior in the low-field region and will gradually change to the linear dependence in the high-field region~\cite{A.V.Stier2018}.
(iii) More importantly, in the Landau quantization space, we reveal that in an exciton state, the dominant free electron-hole pair component may shift with the magnetic field and Coulomb interactions.  We further calculate the phase diagrams of the dominant pair component of the exciton states.  The phase diagrams indicate that in an exciton state $nl$, the magnetic field will drive the dominant component to the electron-hole pair $\{n_e=n+l-1,n_h=n-1\}$, whereas the Coulomb interactions drive it to the pair of the lower index; their competitions would determine the exciton state composition.  
Our theoretical study could provide deeper insights into the exciton states in monolayer TMDs and pave the way for their applications in future electronic devices.

\section{Model}

Consider 2D bound excitons in monolayer WSe$_2$.  Within the effective mass approximation, when the magnetic field is absent, the model Hamiltonian describing the exciton dynamics is written as~\cite{G.H.Wannier} 
\begin{align}
H_0=\frac{\boldsymbol p_e^2}{2m_e}+\frac{\boldsymbol p_h^2}{2m_h}+V(r),    
\end{align}
where $\boldsymbol p_{e(h)}$ denotes the momentum for the electron (hole) and $m_{e(h)}$ is the mass,   
$V(r)$ is the Coulomb interaction potential, and $r=|\boldsymbol r_e-\boldsymbol r_h|$ is the electron-hole separation.  In a 2D system, the nonlocal screening effects will enhance the Coulomb interactions~\cite{N.S.Rytova, Keldysh, P.Cudazzo}, which can be described by the Keldysh potential~\cite{M.Donck2019, D.V.Tuan2018} 
\begin{align} 
V(r)=-\frac{e^2}{8\epsilon_0 r_0}
\Big[H_0\Big(\frac{\epsilon_v r}{r_0}\Big)-Y_0\Big(\frac{\epsilon_v r}{r_0}\Big)\Big].  
\label{Vkr}  
\end{align}
Here $H_0(x)$ and $Y_0(x)$ are the Struve function and Bessel function of the second kind, respectively, $r_0=2\pi\chi_{\text{2D}}$ denotes the screening length characterized by the dielectric nature of monolayer WSe$_2$, with $\chi_{\text{2D}}$ being the 2D polarizability, $\epsilon_0$ is the vacuum dielectric constant,  and $\epsilon_v$ is the relative dielectric constant that is related to the encapsulating hexagonal BN slabs. 
For the large electron-hole separation $r\gg r_0$, the Keldysh potential behaves as a Coulomb-like potential, $V(r)\sim\frac{1}{r}$; whereas for the small separation $r\ll r_0$, it diverges weakly as $V(r)\sim\text{log}(r)$.  Thus the Keldysh potential is expected to induce exciton states that are markedly different from those caused by the hydrogen-like potential.  

When a uniform magnetic field $\boldsymbol B=B\hat{\boldsymbol e}_z$ is applied on the 2D system, we use the Peierls substitution to replace the momenta as $\boldsymbol p\rightarrow\boldsymbol p-q\boldsymbol A$, in which $\boldsymbol A$ is the magnetic vector potential and $q=\mp e$ is the charge for the electron and hole, respectively.  Then the Hamiltonian becomes  
\begin{align}
H=\frac{(\boldsymbol p_e+e\boldsymbol A_e)^2}{2m_e}
+\frac{(\boldsymbol p_h-e\boldsymbol A_h)^2}{2m_h}+V(r).  
\end{align}
We will solve $H$ in both real space and Landau quantization space.  In the following, we set the model parameters as $r_0=5$ nm, $\epsilon_v=3.97$, $m_e=0.29m_0$, $m_h=0.64m_0$, and $m_0$ is the electron mass.  The mass parameters are taken from the DFT calculations~\cite{A.Kormanyos} and are consistent with the magnetoabsorption spectroscopy experiments~\cite{A.V.Stier2018}.  
There also exist some other effective mass values in monolayer WSe$_2$, such as $m_e=0.5m_0$ from the single-electron-transistor experiments~\cite{M.V.Gustafsson} and $m_h=0.45m_0$ from the magnetotransport measurements~\cite{B.Fallahazad}.  We believe that the different effective mass parameters would not change the main conclusions of this paper.

\section{Calculation Methods}

In this section, we present a detailed illustration of the calculation methods, with the real space method following from Ref.~\cite{A.V.Stier2018, A.Spiridonova} and the Landau quantization space method from Ref.~\cite{D.V.Tuan2025}.

\subsection{Real space}

In real space, we switch to the center-of-mass and relative motion coordinates, $\boldsymbol R=\frac{1}{M}(m_e\boldsymbol r_e+m_h\boldsymbol r_h)$ and $\boldsymbol r=\boldsymbol r_e-\boldsymbol r_h$, with $M=m_e+m_h$ being the total mass.  For the Schr\"odinger equation $H\Psi=\varepsilon\Psi$, we perform the gauge transformation to the wavefunction $\Psi\rightarrow e^{if}\Psi$.  
If we choose the gauge $f=\frac{eB}{2\hbar}(x_ey_h-y_ex_h)$~\cite{S.N.Walck}, the Hamiltonian will be transformed as 
$H\rightarrow e^{-if}H e^{if}$ and is given as   
\begin{align}
H=&\frac{\big(\frac{m_{e}}{M}\boldsymbol P+\boldsymbol p+e\boldsymbol A_r\big)^2}{2m_e}
+\frac{\big(\frac{m_h}{M}\boldsymbol P-\boldsymbol p+e{\boldsymbol A}_r\big)^2}{2m_h}+V(r).    
\end{align}
Here $\boldsymbol P$ and $\boldsymbol p$ denote the center-of-mass momentum and relative momentum of the exciton, respectively.   
Focusing on the vanishing $\boldsymbol P$, we obtain the Hamiltonian for the relative motion of the electron-hole pair,  
\begin{align} 
H=\frac{p^2}{2m_r}
+\Big(\frac{1}{m_e}-\frac{1}{m_h}\Big) e\boldsymbol p\cdot\boldsymbol A_r
+\frac{e^2 A^2}{2m_r}+V(r),  
\label{rel-H}
\end{align}
where $m_r^{-1}=m_e^{-1}+m_h^{-1}$ is the reduced mass of the exciton.  Adopting the symmetric gauge $\boldsymbol A_r=\frac{1}{2}B(\hat{\boldsymbol e}_z\times \boldsymbol r)$, we have
$\boldsymbol p\cdot\boldsymbol A_r=\frac{1}{2}B L_z$, with $L_z=-i\hbar\frac{\partial}{\partial\theta}$ denoting the angular momentum operator  the $z$ direction.  
In Eq.~(\ref{rel-H}), the first term gives the kinetic energy of the relative motion, the second term gives the interactions between the exciton magnetic moment and the magnetic field, which can be identified as the Zeeman shift, and the third term represents the weak quadratic confinement potential that determines the diamagnetic shift~\cite{S.N.Walck}. 

Since $H$ owns the radial symmetry, we have the commutation relation $[H,L_z]=0$.  As a result, the wavefunction can be divided into the radial and angular components, $\Psi_{nl}(\boldsymbol r)=R_{nl}(r)Y_l(\theta)$, with $Y_l(\theta)=\frac{1}{\sqrt{2\pi}}e^{il\theta}$.  
Similar to the 2D hydrogen atom~\cite{X.L.Yang}, we use the index $nl$ to label the energy levels of the excitons, in which the principal quantum number $n=1,2,\cdots$, and the magnetic quantum number $l=0,\pm1,\pm2,\pm3,\cdots,\pm(n-1)$, corresponding to the $s,p\pm,d\pm, f\pm, \cdots$ states.  
By inserting $\Psi_{nl}$ into Eq.~(\ref{rel-H}), we obtain the reduced radial equation, 
\begin{align} 
&\Big[-\frac{\hbar^2}{2m_r}\Big(\partial_r^2+\frac{1}{r}\partial_r-\frac{l^2}{r^2}\Big)
+\Big(\frac{1}{m_e}-\frac{1}{m_h}\Big)\frac{\hbar leB}{2}+\frac{e^2 B^2}{8m_r}r^2
\nonumber\\
&+V(r)\Big] R_{nl}(r)=\varepsilon_{nl}R_{nl}(r). 
\label{Seq}
\end{align}

As Eq.~(\ref{Seq}) includes the Keldysh potential $V(r)$, it cannot be solved analytically and we resort to numerics~\cite{M.Molas}.
To numerically solve the second-order differential equation, we introduce the first derivative of the radial wave function 
\begin{align}
D_{nl}(r)=\partial_rR_{nl}(r).  
\label{Deq}
\end{align}
In the calculations, both $R(r)$ and $D(r)$ will be represented on the equidistant grid points $r_i$ from the origin $r_1=0$ to the cutoff $r_N=r_{\text{cut}}$.  Here $r_{\text{cut}}$ should be large enough to indicate the asymptotic behavior at infinity.  On the other hand, the separation between neighboring grids $\delta r=r_{i+1}-r_i$ should be small enough to ensure the accuracy of the results.  In the following, we will set $\delta r=0.05$ nm and $r_{\text{cut}}=60$ nm (see Appendix A).  

We transform the differential equations~(\ref{Seq}) and (\ref{Deq}) into finite-difference equations.  Explicitly, Eq.~(\ref{Seq}) transforms into a set of linear equations:  
\begin{widetext}
\begin{align}
-\frac{\hbar^2}{2m_r}\Big[\frac{D(r_{i+1})-D(r_i)}{r_{i+1}-r_i}+\frac{D(r_{i+1})+D(r_i)}{r_{i+1}+r_i}\Big] 
&+\Big[\frac{\hbar^2l^2}{2m_r(\frac{r_i+r_{i+1}}{2})^2} 
+\frac{e^2 B^2}{8m_r}\Big(\frac{r_{i+1}+r_i}{2}\Big)^2 
+\Big(\frac{1}{m_e}-\frac{1}{m_h}\Big)\frac{\hbar leB}{2}
\nonumber\\
&+V\Big(\frac{r_i+r_{i+1}}{2}\Big)\Big]\times\frac{R(r_{i+1})+R(r_i)}{2}
=\varepsilon\times\frac{R(r_{i+1})+R(r_i)}{2}, 
\label{leq1}
\end{align}
\end{widetext}
and Eq.~(\ref{Deq}) transforms into another set of linear equations, 
\begin{align}
&\frac{1}{2}\big[D(r_{i+1})+D(r_i)\big]=\frac{R(r_{i+1})-R(r_i)}{r_{i+1}-r_i}. 
\label{leq2}
\end{align}
Note that in both Eqs.~(\ref{leq1}) and~(\ref{leq2}), the index $i$ runs from $i=1$ to $i=N-1$. 
To solve the $2N$ unknowns, we consider two boundary conditions: (i) the vanishing wavefunction at $r_N=r_{\text{cut}}$ and (ii) the vanishing derivative at $r_1=0$, which are given as 
\begin{align}
R(r_N)=0,  \quad D(r_1)=0. 
\label{beq}
\end{align}

The $2N-2$ equations in Eqs.~(\ref{leq1}) and~(\ref{leq2}) together with the two boundary condition equations in Eq.~(\ref{beq}) constitute $2N$ linear equations with $2N$ unknowns. 
Since only $N-1$ equations in Eq.~(\ref{leq1}) include the eigenenergy $\varepsilon$, the set of $2N$ equations do not formulate a standard matrix eigenvalue problem, but instead a generalized eigenvalue problem,  
\begin{align}
AV=\varepsilon B V. 
\label{gen}
\end{align}
Here the matrix $A$ incorporates the left-hand sides of Eqs.~(\ref{leq1})-(\ref{beq}), and $V$ is a vector containing the total $2N$ components of $R(r_i)$ and $D(r_i)$ on the grid points.  
In matrix $B$, the elements are mostly zero except for those corresponding to the right-hand side of Eq.~(\ref{leq1}).  In the numerical calculations, we use the function dggev.f in the Lapack library
to solve Eq.~(\ref{gen}).

\subsection{Landau quantization space} 

Consider a free electron or hole in a 2D system with area $L_x\times L_y$.  Under a perpendicular magnetic field $\boldsymbol B=B\hat{\boldsymbol e}_z$, we choose the vector potential in the Landau gauge as $\boldsymbol A=Bx\hat{\boldsymbol e}_y$.  Then the Hamiltonian for the free electron (hole) is written as  
\begin{align}
H_{e(h)}=\frac{1}{2m_{e(h)}}\big[p_{e(h)x}^2+(p_{e(h)y}\pm eBx)^2\big].   
\end{align}
The upper (lower) sign denotes the electron (hole) case.  Since $[H_{e(h)},p_{e(h)y}]=0$, the wave vector $k_{e(h)y}$ remains as a good quantum number.  The eigenenergy and eigenstate of the $n$-th LL are given as 
\begin{align}
&\varepsilon_{n_{e(h)}}=\Big(n_{e(h)}+\frac{1}{2}\Big)\hbar\omega_{e(h)}, 
\label{En}
\\
&\psi_{n_{e(h)},k_{e(h)y}}(x,y)=\frac{e^{ik_{e(h)y}y}}{\sqrt{L_yl_B}}\tilde H_{n_{e(h)}}\Big(\frac{x-x_{e(h)}^*}{l_B}\Big).  
\end{align} 
where $\omega_{e(h)}=\frac{eB}{m_{e(h)}}$ is the cyclotron frequency, $x_{e(h)}^*=\mp k_{e(h)y}l_B^2$ gives the equilibrium position of the oscillations, $l_B=\sqrt{\frac{\hbar}{eB}}$ is the magnetic length, $\tilde H_n(x)$ denotes the Hermite-Gaussian function that is made up of the product of the Hermite polynomial $H_n(x)$ and the Gaussian weight function 
\begin{align}
\tilde H_n(x)=\frac{1}{\sqrt{2^nn!\sqrt\pi}}e^{-x^2/2}H_n(x). 
\end{align}
Note that in Eq.~(\ref{En}), as the LL energy is independent of $k_{e(h)y}$, the LLs are highly degenerate, with the degeneracy determined by $k_{e(h)y}$ as $N_L=\frac{L_xL_y}{2\pi l_B^2}$.  

When transforming to momentum space, the wave function becomes 
\begin{align}
\psi_{n_{e(h)}}(k_{e(h)x},k_{e(h)y})&=(-i)^{n_{e(h)}}\sqrt{\frac{2\pi l_B}{L_x}} e^{\pm ik_{e(h)x} k_{e(h)y}l_B^2}
\nonumber\\
&\times\tilde H_{n_{e(h)}}(k_{e(h)x}l_B), 
\end{align}
where the phase factor $e^{\pm ik_{e(h)x} k_{e(h)y} l_B^2}=e^{-ik_{e(h)x} x_{e(h)}^*}$ comes from the equilibrium position $x_{e(h)}^*$.  In the second-quantization form, we rewrite the wavefunction  as~\cite{D.V.Tuan2025}  
\begin{align}
|n_{e(h)},k_{e(h)y}\rangle&=i^{-n_{e(h)}}\sqrt{\frac{2\pi l_B}{L_x}} 
\sum_{k_x} e^{\pm ik_{e(h)x} k_{e(h)y}l_B^2}
\nonumber\\ 
&\times\tilde H_n(k_{e(h)x}l_B) |\boldsymbol k_{e(h)}\rangle,  
\end{align}
where $|\boldsymbol k_{e(h)}\rangle=c_{\boldsymbol k_{e(h)}}^\dagger|0\rangle$.  Note that $|n_{e(h)},k_{e(h)y}\rangle$ constitutes the basis of the Landau quantization space.  

We consider the direct interactions between the two particles 1 and 2.  In the basis $|n_{e(h)},k_{e(h)y}\rangle$, the matrix element for the interaction potential is given as $V_{n_1,k_{y1};n_2,k_{y2}}^{n_1',k_{y1}';n_2',k_{y2}'}
=\langle n_1',k_{y1}';n_2',k_{y2}'|V|n_1,k_{y1};n_2,k_{y2}\rangle$.  Here the index without and with prime denote the initial and final states, respectively.  After a direct calculation, we obtain~\cite{D.V.Tuan2025}  
\begin{align}
V_{n_1,k_{y1};n_2,k_{y2}}^{n_1',k_{y1}';n_2',k_{y2}'} 
&=\sum_{\boldsymbol q}e^{\mp iq_xk_{y1}l_B^2\pm iq_xk_{y2}l_B^2}
S_{n_1}^{n_1'}(\boldsymbol q) S_{n_2}^{n_2'}(-\boldsymbol q) 
\nonumber\\
&\times V(q) \delta_{k_{y1}',k_{y1}+q_y}\delta_{k_{y2}',k_{y2}-q_y}. 
\label{Vmatrix}
\end{align}
Similarly, the upper (lower) sign denotes the electron (hole) case.  
Here $V(q)$ is the interaction potential in momentum space, the two $\delta$-functions indicate that the momentum is conserved in the scattering process, and $S_n^{n'}(\boldsymbol q)$ is the form factor given by 
\begin{align}
S_n^{n'}(\boldsymbol q)&=i^{\Delta n} e^{\mp iq_xq_yl_B^2} \frac{2\pi l_B}{L_x}
\sum_{k_x} e^{\mp ik_xq_yl_B^2} 
\nonumber\\
&\times\tilde H_n(k_xl_B) \tilde H_{n'}(k_xl_B+q_xl_B),  
\end{align}
with the index difference $\Delta n=n-n'$.  
To calculate $S_n^{n'}(\boldsymbol q)$, we can transform the summation over $k_x$ into integration and further complete the integration with the help of the generating function of the Hermite polynomials.  Then we have 
\begin{align}
S_n^{n'}(\boldsymbol q)&=i^{\Delta n} \sqrt{\frac{n!n'!}{2^{n+n'}}} e^{\mp\frac{i}{2}\alpha\beta}  e^{-\frac{\alpha^2+\beta^2}{4}}\sum_{m=0}^{\text{min}(n,n')} \frac{2^m}{m!} 
\nonumber\\
&\times\frac{(-\alpha+i\beta)^{n-m}}{(n-m)!}
\frac{(\alpha+i\beta)^{n'-m}}{(n'-m)!}, 
\label{Sn}
\end{align} 
where the dimensionless parameters are $\alpha=q_xl_B$ and $\beta=q_yl_B$.  
Similar expressions in Eqs.~(\ref{Vmatrix})-(\ref{Sn}) of the interactions between particles under a magnetic field can also be found in several previous studies~\cite{C.H.Zhang2007, C.H.Zhang2008, R.Cote, M.O.Goerbig}.  

To simplify Eq.~(\ref{Sn}), we define the orthonormal function $\tilde L_n^{n'}(x)$ as 
\begin{align}
\tilde L_n^{n'}(x)=\sqrt{\frac{n!}{(n+n')!}x^{n'}e^{-x}} L_n^{n'}(x), 
\end{align} 
where $L_n^{n'}(x)$ is the generalized Laguerre polynomial~\cite{I.S.Gradshteyn}.  Then $S_n^{n'}(\boldsymbol q)$ is rewritten as
\begin{align}
S_n^{n'}(\boldsymbol q)=i^{|\Delta n|} 
e^{\mp\frac{i}{2}q_xq_yl_B^2\mp i\Delta n\theta} 
\tilde L_{\tilde n}^{|\Delta n|}\Big(\frac{q^2 l_B^2}{2}\Big), 
\end{align} 
where $\tilde n$=min$\{n,n'\}$ is the minimum between the initial state $n$ and final state $n'$.  
In the numerical calculations, $\tilde L_n^{n'}(x)$ does not diverge even when the index $n$ and $n'$ are much large, thus, $\tilde L_n^{n'}(x)$ is more controllable and favorable than $L_n^{n'}(x)$.  

Without the Coulomb interactions, the Schr\"odinger equation for a free electron-hole pair is 
\begin{align}
(H_e+H_h)|\Psi_0\rangle=(\varepsilon_{n_e}+\varepsilon_{n_h})|\Psi_0\rangle,  
\end{align} 
where the free electron-hole pair wave function is $|\Psi_0\rangle=|n_e,k_e;n_h,k_h\rangle=|n_e,k_e;n_h,K-k_e\rangle$, 
and $K=k_e+k_h$ is the center-of-mass wave vector in the $y$ direction.  Note that $k_{e(h)}=k_{e(h)y}$.  
When turning on the Coulomb interactions between the electrons and holes, we project the bound exciton states onto the space spanned by the free electron-hole pair states, which is the kernel of solving the present interaction problem.  This means that the bound exciton state will be expressed as a superposition of the free pair states, 
\begin{align}
|\Psi_K\rangle=\sum_{n_e,k_e;n_h}\phi_K(n_e,k_e;n_h)|\Psi_0\rangle, 
\end{align}
where $\phi_K(n_e,k_e;n_h)$ is the expansion coefficient.  Since the wave vector $K$ remains unchanged to the Coulomb interaction, it is conserved and gives a constant of motion.  
It is interesting to see that as the separation between the equilibrium positions of the electron and hole is $x_h^*-x_e^*=Kl_B^2$, the separation also remains unchanged.  Then the Schr\"odinger equation including the Coulomb interactions is written as  
\begin{align}
\big[H_e+H_h+V\big(|\boldsymbol r_e-\boldsymbol r_h|\big)\big]|\Psi_K\rangle =\varepsilon_K|\Psi_K\rangle, 
\label{Seq-in}
\end{align}
from which the expanding coefficient $\phi_K$ can be solved.  

We left-multiply Eq.~(\ref{Seq-in}) by the state $\langle\Psi_0'|=\langle n_e',k_e';n_h',K-k_e'|$ and arrive at 
\begin{align}
&\sum_{n_e,n_h}\big[(\varepsilon_{n_e}+\varepsilon_{n_h})\delta_{n_e,n_e'}\delta_{n_h,n_h'} 
+\tilde V_{n_e,n_h}^{n_e',n_h'}\big]\phi_K(n_e,n_h)
\nonumber\\
&=\varepsilon_K \phi_K(n_e',n_h'), 
\label{Seq2}
\end{align}
where the contracted coefficient is $\phi_K(n_e,n_h)=\sum_{k_e}\phi_K(n_e,k_e;n_h)$, and the matrix element for the interaction potential $\tilde V_{n_e;n_h}^{n_e';n_h'}=\sum_{\boldsymbol q}e^{-iKq_xl_B^2} V(q) 
S_{n_e}^{n_e'}(\boldsymbol q) S_{n_h}^{n_h'}(-\boldsymbol q)$ is 
\begin{align}
\tilde V_{n_e,n_h}^{n_e',n_h'} 
&=i^{|\Delta n_e|-|\Delta n_h|}\sum_{\boldsymbol q}e^{-iKq_xl_B^2} V(q) 
e^{-i(\Delta n_e-\Delta n_h)\theta}   
\nonumber\\
&\times\tilde L_{\tilde n_e}^{|\Delta n_e|}\Big(\frac{q^2 l_B^2}{2}\Big)
\tilde L_{\tilde n_h}^{|\Delta n_h|}\Big(\frac{q^2 l_B^2}{2}\Big).  
\label{tildeV}
\end{align}
Note that the matrix element $\tilde V$ in Eq.~(\ref{tildeV}) is different from $V$ in Eq.~(\ref{Vmatrix}), as in the former, the $\delta$-functions related to the wave vectors have been summed over in the calculations and thus do not appear. 
For the Keldysh potential $V(q)$ in momentum space, it can be obtained through the Fourier transformation of $V(r)$, which is given as~\cite{A.Chernikov}  
\begin{align}
V(q)=-\frac{2\pi e^2}{q\epsilon(q)}, 
\end{align}
with the effective dielectric constant $\epsilon(q)=4\pi\epsilon_0(\epsilon_v+r_0q)$.  

We focus on the zero center-of-mass momentum exciton, $K=0$.  When turning the summation over the wave vector  $\boldsymbol q$ into integration, the potential energy matrix element $\tilde V_{n_e,n_h}^{n_e',n_h'}$ becomes
\begin{align} 
&\tilde V_{n_e,n_h}^{n_e',n_h'}
=\frac{1}{4\pi^2}i^{|\Delta n_e|-|\Delta n_h|}
\int_0^\infty qdq  V(q)
\nonumber\\
&\quad\times\tilde L_{\tilde n_e}^{|\Delta n_e|}\Big(\frac{q^2 l_B^2}{2}\Big)
\tilde L_{\tilde n_h}^{|\Delta n_h|}\Big(\frac{q^2 l_B^2}{2}\Big)
\int_0^{2\pi} d\theta e^{-i(\Delta n_e-\Delta n_h)\theta}.  
\end{align}
After completing the integration over the angle $\theta$, the nonvanishing matrix element $\tilde V$ requires the condition 
\begin{align}
\Delta n_e=\Delta n_h, 
\end{align}
which gives the selection rules for the bound excitons~\cite{D.V.Tuan2025}.   
Here, as the selection rules include both the states before and after scatterings, they are different from those in the LL transitions of the noninteracting Dirac electrons/holes~\cite{C.J.Tabert, A.Akrap, E.Martino, W.Duan, Y.X.Wang2022, Y.X.Wang2024}.   

If we define $l_k=n_e-n_h$ and $l_k'=n_e'-n_h'$, the above selection rules indicate that $l_k=l_k'$.  As the potential energy matrix element vanishes for the scatterings between the electron-hole pairs of different $l_k$, i.e., $\tilde V_{i+l_k,i}^{i'+l_k',i'}=0$ for $l_k\neq l_k'$, the Hamiltonian matrix can be partitioned into the block-diagonal form.  Each block includes the electron-hole pairs $(n_e=i+l_k, n_h=i)$, or $(n_e=i, n_h=i+l_k)$ for $-l_k$.  Finally, we arrive at the matrix eigenvalue equation for the bound exciton with the index $l_k$ as 
\begin{align}
(T_{l_k}+{\tilde V}_{l_k})\Psi_{nl_k}=\varepsilon_{nl_k}\Psi_{nl_k}. 
\label{mateq}
\end{align} 
Here the kinetic energy and potential energy matrix elements are given as 
\begin{align}
&T_{l_k}^{i,i'}=\Big[\hbar\omega_e\Big(i+l_k+\frac{1}{2}\Big)
+\hbar\omega_h\Big(i+\frac{1}{2}\Big)\Big]\delta_{i,i'},
\label{Tmat}
\\
&\tilde V_{l_k}^{i,i'}=-\frac{e^2}{l_B}\int_0^\infty \frac{dx}{\epsilon(x)} 
\tilde L_{\tilde i+l_k}^{|\Delta i|}\Big(\frac{x^2}{2}\Big)
\tilde L_{\tilde i}^{|\Delta i|}\Big(\frac{x^2}{2}\Big),  
\label{tildeVmat}
\end{align}
where $\Delta i=i'-i$, $\tilde i=\text{min}\{i,i'\}$, and the dimensionless quantity $x=ql_B$.  The wave function $\Psi_{nl_k}=(\phi_{nl_k}^0,\phi_{nl_k}^1,\cdots)^T$, with the component  $\phi_{nl_k}^i=\phi_{K=0}(i+l_k,i)$.  

In the numerical calculations, completing the integration in Eq.~(\ref{tildeVmat}) is more time-consuming than the matrix diagonalization in Eq.~(\ref{mateq}).  This is because the function $\tilde L_i^{i'}(x)$ oscillates quickly with $x$ when the indices $i$ and $i'$ increase.  Moreover, as the Hilbert space spanned by the electron-hole pairs is infinite, we need to truncate the matrix in Eq.~(\ref{mateq}) and set the dimension cutoff as $N_{\text{cut}}=300$ (see Appendix A).

\section{Main Results}

\subsection{Energy spectrum}

\begin{figure}
	\includegraphics[width=9.2cm]{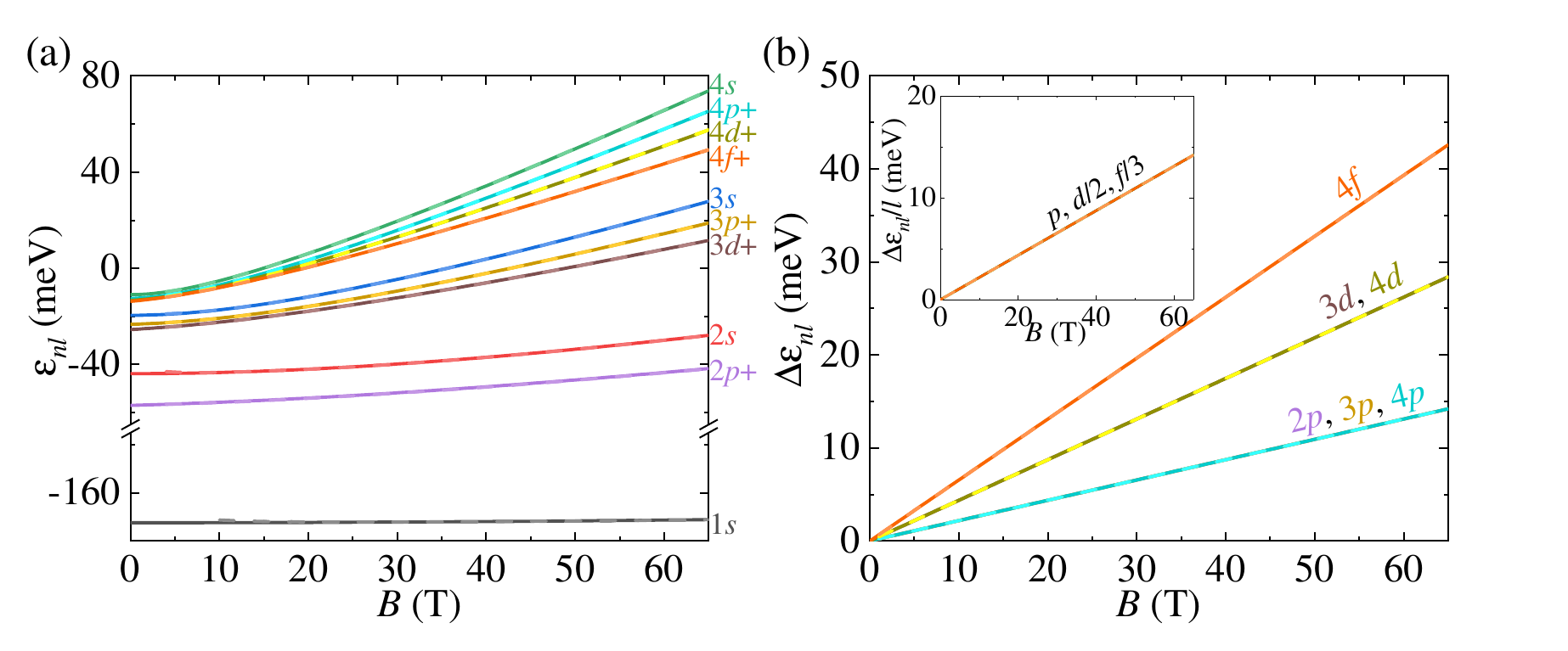}
	\caption{(Color online) (a) The exciton energy spectrum $\varepsilon_{nl}$ vs the magnetic field $B$.  (b) The energy difference $\Delta\varepsilon_{nl}=\varepsilon_{n,+l}-\varepsilon_{n,-l}$ vs $B$.  In both figures, the solid and dashed lines denote the results from real space and Landau quantization space, respectively.  The inset in (b) shows $\Delta\varepsilon_{nl}$ scaled by $l$, all of which collapse on the same line.  Note that for the Landau quantization results, in (a), $\varepsilon_{1s}$ is plotted for $B\geq10$ T and other $\varepsilon_{nl}$ are plotted for $B\geq4$ T, while in (b), all $\Delta\varepsilon_{nl}$ are plotted for $B$ down to zero. }  
	\label{Fig1}
\end{figure} 

First of all, we calculate the exciton energy spectrum $\varepsilon_{nl}$ of monolayer WSe$_2$ in real space and Landau quantization space, and plot $\varepsilon_{nl}$ as a function of the magnetic field $B$ in Fig.~\ref{Fig1}(a).   
Here, the results obtained from the two spaces are represented by the solid and dashed lines, respectively.  Comparing $\varepsilon_{nl}$ in the two spaces, we find the results in the Landau quantization space agree well with those in real space.  
Note that in the Landau quantization space, we have made certain truncations to the magnetic field, with $\varepsilon_{1s}$ plotted for $B\geq10$ T and other levels for $B\geq4$ T.  This is because in the Landau quantization space, the lower energy levels under weak magnetic fields will deviate from those in real space.  Such deviations can be attributed to the finite matrix dimension cutoff (see Appendix A). 

With no magnetic field, the energies of the $ns$ states from real space are given as $\varepsilon_{1s}=-172.44$ meV, $\varepsilon_{2s}=-43.82$ meV, $\varepsilon_{3s}=-19.52$ meV, and $\varepsilon_{4s}=-10.96$ meV.  
Due to the Keldysh potential, these energy levels exhibit a markedly non-hydrogenic Rydberg series and thus cannot be described within a hydrogen-like model.  More importantly, these energy levels are consistent with the previous experimental and theoretical results on the exciton Rydberg states in monolayer WSe$_2$~\cite{A.V.Stier2018, E.Liu2019, A.Spiridonova}, demonstrating the validity of our calculations.   
For the magnetic field over a large range $B\sim(0,65\text{ T})$, the energy levels are grouped for the same $n$.  
With increasing $B$, the ground state $1s$ shows minor blueshift and the excited states exhibit evident blueshifts, which become more apparent for the higher excited states.  
The blueshifts of the exciton states under magnetic fields are in good agreement with the experiments~\cite{A.V.Stier2016a, A.V.Stier2018, E.Liu2019}.  

Consider the energy difference between the exciton states of positive and negative $l(l_k)$,  $\Delta\varepsilon_{nl(l_k)}=\varepsilon_{n,+l(l_k)}-\varepsilon_{n,-l(l_k)}$.  
The results of $\Delta\varepsilon_{nl(l_k)}$ are displayed in Fig.~\ref{Fig1}(b), with $l(l_k)$ growing from $l(l_k)=1$ to $3$.  Note that in the Landau quantization space, all $\Delta\varepsilon_{nl}$ are plotted for the magnetic field down to zero.  
We observe that the lines of $\Delta\varepsilon_{nl}$ pass through the origin and overlap completely with those of $\Delta\varepsilon_{nl_k}$, indicating that $\Delta\varepsilon_{nl}=\Delta\varepsilon_{nl_k}$.  

\begin{figure}
	\includegraphics[width=9.2cm]{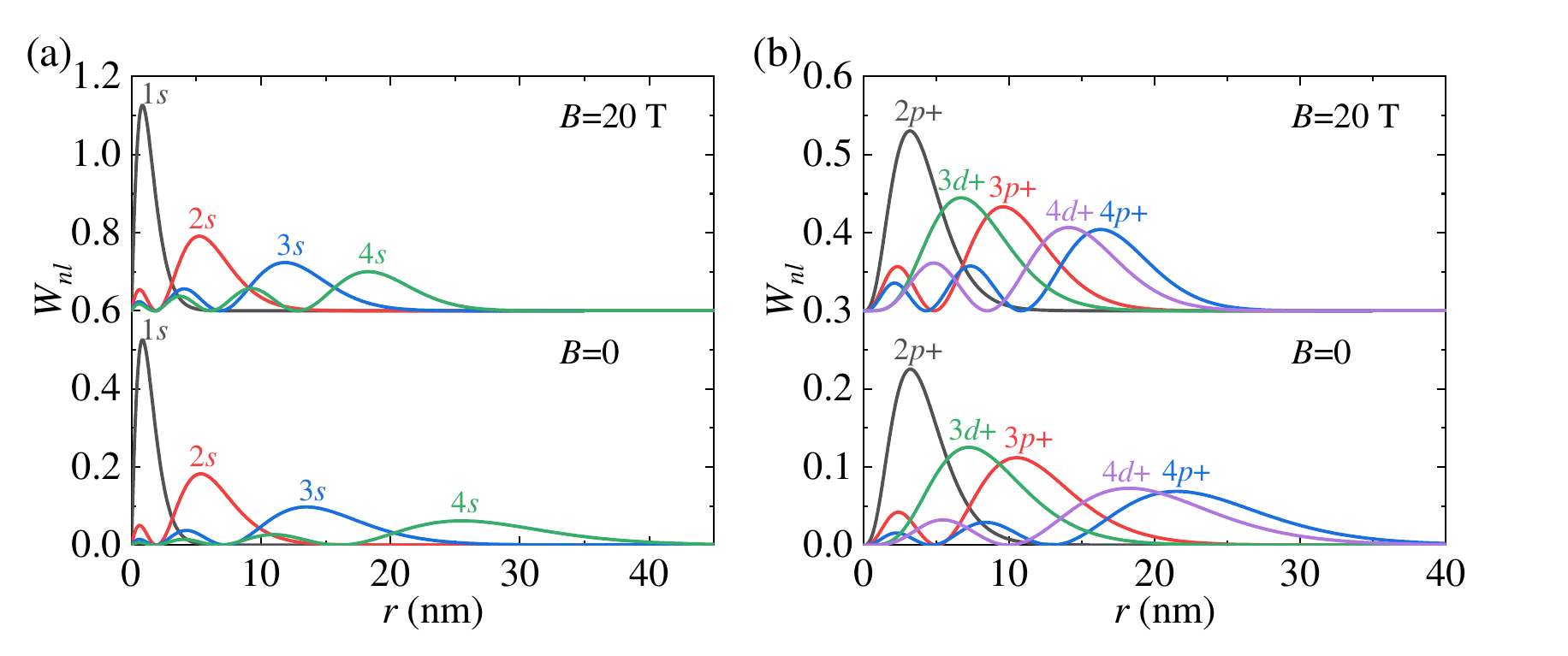}
	\caption{(Color online) The probability $W_{nl}$ of finding the bound exciton vs the electron-hole separation $r$ from the real space calculations, with $l=0$ in (a) and $l=1,2$ in (b).  In each lower and upper figure, the magnetic field is set as $B=0$ and $B=20$ T, respectively.  For clarity, the probability curves under $B=20$ T are shifted vertically by $0.6$ in (a) and $0.3$ in (b).} 
	\label{Fig2}
\end{figure} 

In real space, the energy levels are solved from Eq.~(\ref{Seq}).  When $l$ changes to $-l$, we see that only the second term of the left-hand side will change and the other terms remain unchanged.  As a result, the energy difference is  
\begin{align} 
\Delta\varepsilon_{nl}(B)=\hbar(\omega_e-\omega_h)l. 
\label{Dl}
\end{align} 
Note that when $l$ changes to $-l$ and $B$ remains unchanged, the energy levels are equal to those when $B$ changes to $-B$ and $l$ remains unchanged, $\varepsilon_{n,-l}(B)=\varepsilon_{n,l}(-B)$. 
Thus, the energy difference is $\Delta\varepsilon_{nl}=2l\mu_B^{\text{ex}}B$, which is twice the Zeeman shift and originates from the orbital magnetic moment.  The corresponding Bohr magneton is given as $\mu_B^{\text{ex}}=\frac{e\hbar}{2}(\frac{1}{m_e}-\frac{1}{m_h})$.  Clearly, $\mu_B^{\text{ex}}$ is caused by unequal electron and hole masses and vanishes when $m_e=m_h$.  

On the other hand, in the Landau quantization space, the state $|n,-l_k\rangle$ is obtained from the state $|n,l_k\rangle$ by making the electron-hole transformation $n_e\leftrightarrow n_h$.  Under this transformation, the potential energy matrix element $\tilde V_{l_k}^{i,i'}$ in Eq.~(\ref{tildeVmat}) remains unchanged, but the kinetic energy matrix element $T_{l_k}^{i,i'}$ in Eq.~(\ref{Tmat}) will change, leading to the energy difference  
\begin{align} 
\Delta\varepsilon_{nl_k}
=&\hbar\Big[\omega_e\Big(n+l_k+\frac{1}{2}\Big)+\omega_h\Big(n+\frac{1}{2}\Big)\Big]
\nonumber\\
&-\hbar\Big[\omega_e\Big(n+\frac{1}{2}\Big)+\omega_h\Big(n+l_k+\frac{1}{2}\Big)\Big]
\nonumber\\
=&\hbar(\omega_e-\omega_h)l_k. 
\label{Dlk} 
\end{align}
Combining the observation $\Delta\varepsilon_{nl}=\Delta\varepsilon_{nl_k}$ with Eqs.~(\ref{Dl}) and~(\ref{Dlk}), we have $l=l_k$.  This means that although $l_k$ originates from the selection rules of the electron-hole LL pairing in the Landau quantization space, $n_e=n_h+l_k$, and $l$ denotes the magnetic quantum number in the real space, they are equivalent to each other in labeling the exciton energy levels~\cite{D.V.Tuan2025}.  For simplicity, we drop the subindex $k$ in the following.  
The above analysis of $\Delta\varepsilon_{nl}$ indicates that the energy deviations of the lower levels under weak magnetic field in the Landau quantization space are not caused by the kinetic energy of the electron and hole cyclotron motion, but by the interaction potential energy. 

\begin{table*}[t]  
	\caption{The diamagnetic coefficient $\sigma_{nl}$ and rms radius $r_{nl}$, which are in units of $\mu$eV/T$^2$ and nm, respectively.  The calculation results obtained from Eq.~(\ref{diam1}) as well as the fitting results in Figs.~\ref{Fig3}(b) and~\ref{Fig3}(c) are listed.  For comparison, the experimental data in Ref.~\cite{A.V.Stier2018} and Ref.~\cite{E.Liu2019} are given in the last column.  The fitting slope $k_{nl}$ in unit of meV/T is also listed.}  
	\begin{tabular}{c|cc|ccc|cccc}\hline
		exciton state& \multicolumn{2}{|c|}{results through Eq.~(\ref{diam1})}& 
		\multicolumn{3}{|c|}{fitting results in Figs.~\ref{Fig3}(b) and (c)}& 
		\multicolumn{4}{|c|}{experimental data} 
		\\
		$nl$& $\sigma_{nl}$& $r_{nl}$& $\sigma_{nl}$& $r_{nl}$& $k_{nl}$& $\sigma_{nl}$~\cite{A.V.Stier2018}& $r_{nl}$~\cite{A.V.Stier2018}&   
		$\sigma_{nl}$~\cite{E.Liu2019}& $r_{nl}$~\cite{E.Liu2019}\\
		\hline
		$1s$& $0.31$& $1.68$& $0.31$& $1.68$& $0.032$& $0.31\pm0.02$& $1.7\pm0.1$& $0.24\pm0.1$& $1.6\pm0.4$
		\\
		$2s$& $4.87$& $6.66$& $4.77$& $6.59$& $0.37$& $4.6\pm0.2$& $6.6$& $6.4\pm0.2$& $8.24\pm0.13$   
		\\
		$3s$& $24.2$& $14.85$& $22.73$& $14.39$& $0.96$& $22\pm2$& $14.3\pm1.5$& $27.3\pm1.3$& $17.0\pm0.4$ 
		\\    
		$4s$& $76.2$& $26.34$& $70.23$& $25.29$& $1.58$& /& /& $53.7\pm3.0$& $27.8\pm0.7$ 
		\\
		\hline   
		$2p$& $2.3$& $4.58$& $2.27$& $4.55$& $0.19$& /& /& /& /
		\\
		$3p$& $15.7$& $11.96$& $14.93$& $11.66$& $0.73$& /& /& /& /
		\\
		$4p$& $56.2$& $22.63$& $50.7$& $21.49$& $1.34$& /& /& /& /
		\\  
		\hline  
		$3d$& $9.43$& $9.27$& $9.12$& $9.11$& $0.49$& /& /& /& /
		\\ 
		$4d$& $42.4$& $19.65$& $38.76$& $18.79$& $1.08$& /& /& /& /
		\\
		\hline 
		$4f$& $28.4$& $16.08$& $26.3$& $15.48$& $0.81$& /& /& /& / 
		\\  
		\hline
	\end{tabular}		
	\label{table1}
\end{table*}

\begin{figure*}
	\includegraphics[width=18cm]{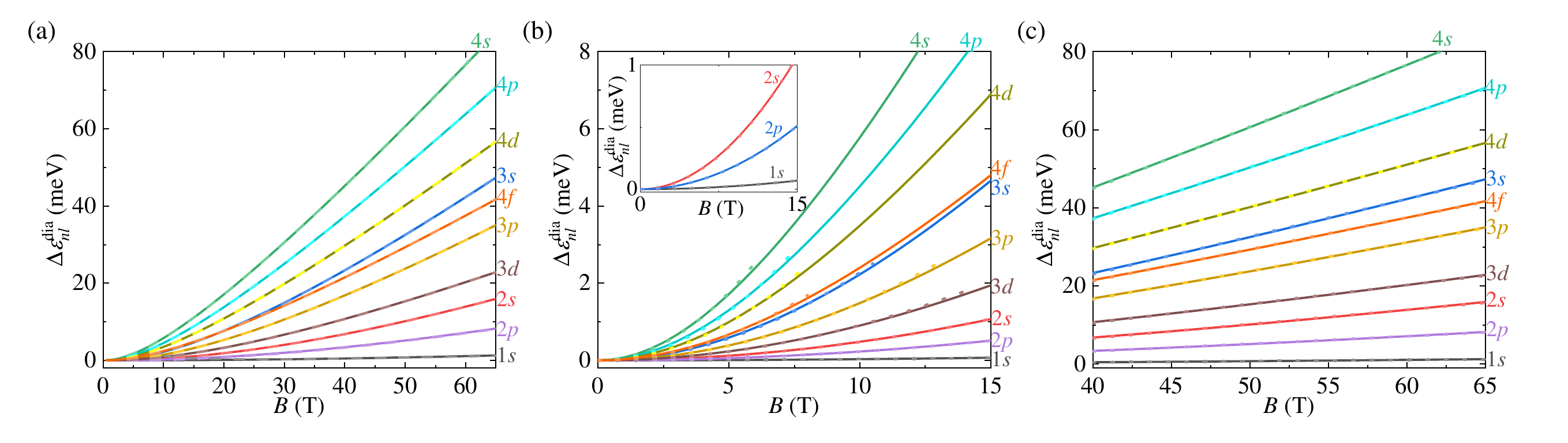}
	\caption{(Color online) (a) The diamagnetic shift $\Delta\varepsilon_{nl}^{\text{dia}}$ obtained from Eq.~(\ref{diam2}) vs the magnetic field $B$ up to 65 T.  The solid and dashed lines denote the results from real space and Landau quantization space, respectively.  Note that for the Landau quantization results, $\Delta\varepsilon_{1s}^{\text{dia}}$ is plotted for $B\geq10$ T and other $\Delta\varepsilon_{nl}^{\text{dia}}$ are plotted for $B\geq4$ T, with the truncations the same as Fig.~\ref{Fig1}(a). 
	(b) The low-field $\Delta\varepsilon_{nl}^{\text{dia}}$ is fitted through Eq.~(\ref{diam}), from which the diamagnetic coefficient $\sigma_{nl}$ and rms radius $r_{nl}$ are obtained and listed in Table~\ref{table1}.  
	(c) The high-field $\Delta\varepsilon_{nl}^{\text{dia}}$ is fitted by using Eq.~({\ref{lin}}), from which the slope $k_{nl}$ is obtained and also listed in Table~\ref{table1}.  In both (b) and (c), the solid and dotted lines denote the real space results and the fitted results, respectively.  The inset of (b) gives an enlarged plot for the three lowest levels, $1s$, $2p$ and $2s$. }  
	\label{Fig3}  
\end{figure*} 

Moreover, in Fig.~\ref{Fig1}(b), we observe that $\Delta\varepsilon_{nl}$ of the same $l$ but different $n$ lie on the same line.  
When scaling $\Delta\varepsilon_{nl}$ by $l$, they all coincide with $\Delta\varepsilon_{n's}$, as seen in Fig.~\ref{Fig1}(b) inset.  These observations can be easily understood from Eqs.~(\ref{Dl}) or~(\ref{Dlk}) that $\Delta\varepsilon_{nl}$ does not depend on $n$, but on $l$ as $\Delta\varepsilon_{nl}=l\Delta\varepsilon_{n's}$.  

Besides the energy spectrum, the real space wavefunction $\Psi_{nl}(\boldsymbol r)=R_{nl}(r)Y_l(\theta)$ can also be obtained.  According to $\Psi_{nl}(\boldsymbol r)$, the probability $W_{nl}(r)dr$ of finding the bound exciton with the electron-hole separation in the range $(r,r+dr)$ is written as   
\begin{align}
W_{nl}(r)dr=\int_0^{2\pi} d\theta dr rR_{nl}^2(r)Y_l^2(\theta)=R_{nl}^2(r)rdr.  
\end{align} 
The results of $W_{nl}$ are displayed in Fig.~\ref{Fig2} as a function of $r$.  
In Figs.~\ref{Fig2}(a) and \ref{Fig2}(b) when $B=0$, we see that (i) for the ground state, $W_{1s}$ has no node, while for the excited states, $W_{nl}$ owns $n-l-1$ nodes (except $r=0$ and $r=\infty$);  
(ii) for the state $nl$, $W_{nl}$ will extend to larger $r$ when $n$ increases (or $l$ decreases), meaning that the excitons are loosely bound.   
These results are the same as those in 2D hydrogen atoms~\cite{X.L.Yang}.  
At a finite magnetic field $B=20$ T, the above conclusions do not change, but $W_{nl}$ will be compressed to smaller $r$, suggesting that the excitons become more tightly bound.

\subsection{Diamagnetic shift}

\begin{figure*}
	\includegraphics[width=18cm]{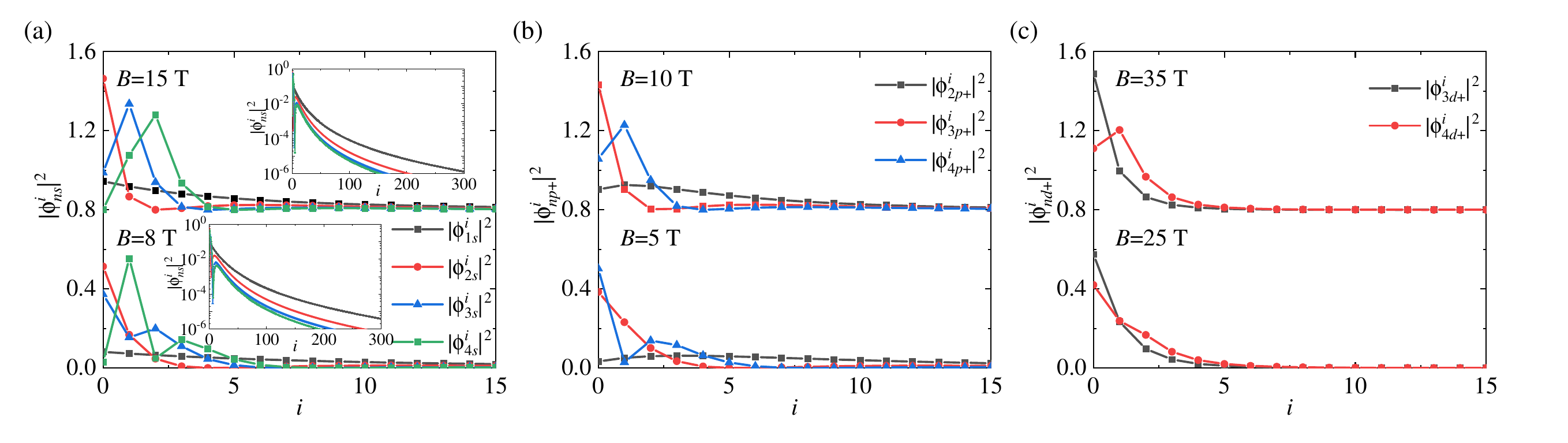}
	\caption{(Color online) The weight $|\phi_{nl}^i|^2$ of the free electron-hole pair $\{n_e=i+l,n_h=i\}$ in the exciton state $nl$ vs the hole LL index $i$ from the Landau quantization space calculations.  In (a)-(c), the weights are shown for the $s$, $p+$ and $d+$ states, respectively.  The lower and upper figures are the results under different magnetic fields.  
	For clarity, the weights in the upper figures are shifted vertically by $0.8$.  The insets in (a) show the weight $|\phi_{ns}^i|^2$ vs the index $i$ up to $i=300$.}  
	\label{Fig4}  
\end{figure*}

In this section, we study the diamagnetic shift of the bound excitons.  
The diamagnetic shift is expected to increase quadratically with the magnetic field~\cite{R.S.Knox, N.Miura, S.N.Walck} and is taken as a measure of the confinement effect~\cite{T.Someya}.  
For the state $nl$, the quadratic diamagnetic shift is expressed as~\cite{N.Miura, R.S.Knox}
\begin{align} 
\Delta\varepsilon_{nl}^{\text{dia}}(B)=\sigma_{nl}B^2,  
\label{diam}
\end{align}
where $\sigma_{nl}$ denotes the diamagnetic coefficient.  The diamagnetic coefficient can determine the exciton effective mass, radius and the dielectric properties of a material~\cite{A.V.Stier2016a, A.V.Stier2016b}.  

In Eq.~(\ref{Seq}), since only the third term of the left-hand side exhibits the quadratic field dependence, we obtain the diamagnetic shift as  
\begin{align}
\Delta\varepsilon_{nl}^{\text{dia}}(B)=\frac{e^2}{8m_r}\langle r_{nl}^2\rangle B^2, 
\label{diam1} 
\end{align}
where the expectation value $\langle r_{nl}^2\rangle$ is with respect to the zero-field  eigenstate~\cite{S.N.Walck}.  The rms radius is related to $\sigma_{nl}$ as $r_{nl}=\sqrt{\langle r_{nl}^2\rangle}=\frac{\sqrt{8m_r\sigma_{nl}}}{e}$ and can be used to estimate the spatial extent of the exciton wavefunctions.  
By using the zero-field wavefunction $\Psi_{nl}(\boldsymbol r)$, we calculate $\langle r_{nl}^2\rangle$ as
\begin{align}
\langle r_{nl}^2\rangle=\int_0^\infty rdr\int_0^{2\pi}d\theta r^2R_{nl}^2(r) Y_l^2(\theta)
=\int_0^\infty dr r^3R_{nl}^2(r). 
\label{rnl} 
\end{align}
From $\langle r_{nl}^2\rangle$, the diamagnetic coefficient $\sigma_{nl}$ and rms radius $r_{nl}$ can be calculated, with the results listed in Table~\ref{table1}.  We see that both $\sigma_{nl}$ and $r_{nl}$  increase for the higher exciton states, reflecting the stronger diamagnetic behavior and larger rms radius.   

Another route to calculate the diamagnetic shift is from the average energy of the $nl$ and $n,-l$ states subtracting the zero-field energy~\cite{D.V.Tuan2025}, 
\begin{align}
\Delta \varepsilon_{nl}^{\text{dia}}(B)
=\frac{1}{2}\big[\varepsilon_{nl}(B)+\varepsilon_{n,-l}(B)\big]-\varepsilon_{n,l}(B=0).  
\label{diam2} 
\end{align} 
Here the average energy cancels out the interactions between the exciton magnetic moment and the magnetic field, and the zero-field energy subtracts the contributions from the kinetic energy of relative motion and the Keldysh potential energy, as seen in Eq.~(\ref{Seq}).  
Following this way, we calculate $\Delta\varepsilon_{nl}^{\text{dia}}$ and plot the results in Fig.~\ref{Fig3}(a), where the solid and dashed lines denote the results in the real space and Landau quantization space, respectively.  We see that the results in the two spaces agree well with each other.  Note that in the Landau quantization space, the magnetic field truncations are the same as Fig.~\ref{Fig1}(a).  
 
In the low-field region, we fit $\Delta\varepsilon_{nl}^{\text{dia}}$ by using Eq.~(\ref{diam}) and the fitting results are displayed by the dotted lines in Fig.~\ref{Fig3}(b).  
To see the fittings of the lowest three levels $1s$, $2p$ and $2s$ more clearly, an enlarged plot is given in Fig.~\ref{Fig3}(b) inset.  
According to the fittings, we obtain $\sigma_{nl}$ and $r_{nl}$, which are listed in Table~\ref{table1}. 

In Table~\ref{table1}, we also list the experimental data for the $ns$ exciton states in monolayer WSe$_2$~\cite{A.V.Stier2018, E.Liu2019}.  We find that the two theoretical results from Eq.~(\ref{diam1}) and the fittings agree well with the experimental data, demonstrating the validity of the effective mass exciton model as well as the chosen parameters.  In addition, we note that $\sigma_{1s}$ and $r_{1s}$ are close to those reported in monolayer WS$_2$ experiment~\cite{A.V.Stier2016b}. 

On the other hand, in the high-field region, $\Delta\varepsilon_{nl}^{\text{dia}}$ exhibits a linear dependence.  Thus, in the range $B\sim(40,65)$T, we fit the diamagnetic shift by using the linear relationship  
\begin{align}
\Delta\varepsilon_{nl}^{\text{dia}}(B)=k_{nl}B+b_{nl}, 
\label{lin}
\end{align}
where $k_{nl}$ and $b_{nl}$ denote the slope and intercept of the $ns$ state, respectively.  
The fitting results are displayed by the dotted lines in Fig.~\ref{Fig3}(c) and the obtained $k_{nl}$ are listed in Table~\ref{table1}.  
We see that $k_{nl}$ generally increases with the exciton state (except for the $3s$ and $4f$ states, whose orders are reversed), meaning that $\Delta\varepsilon_{nl}^{\text{dia}}$ grows more quickly with the magnetic field for the higher exciton states.  
These results demonstrate that the quadratic diamagnetic shift represents a low-field behavior~\cite{S.N.Walck}; with increasing magnetic field, $\Delta\varepsilon_{nl}^{\text{dia}}$ will gradually change from the low-field quadratic dependence $\Delta\varepsilon_{nl}\sim B^2$ to the high-field linear behavior $\Delta\varepsilon_{nl}\sim B$, as was observed in the experiment~\cite{A.V.Stier2018}.

\subsection{Exciton state composition} 

\begin{figure*}
	\includegraphics[width=18cm]{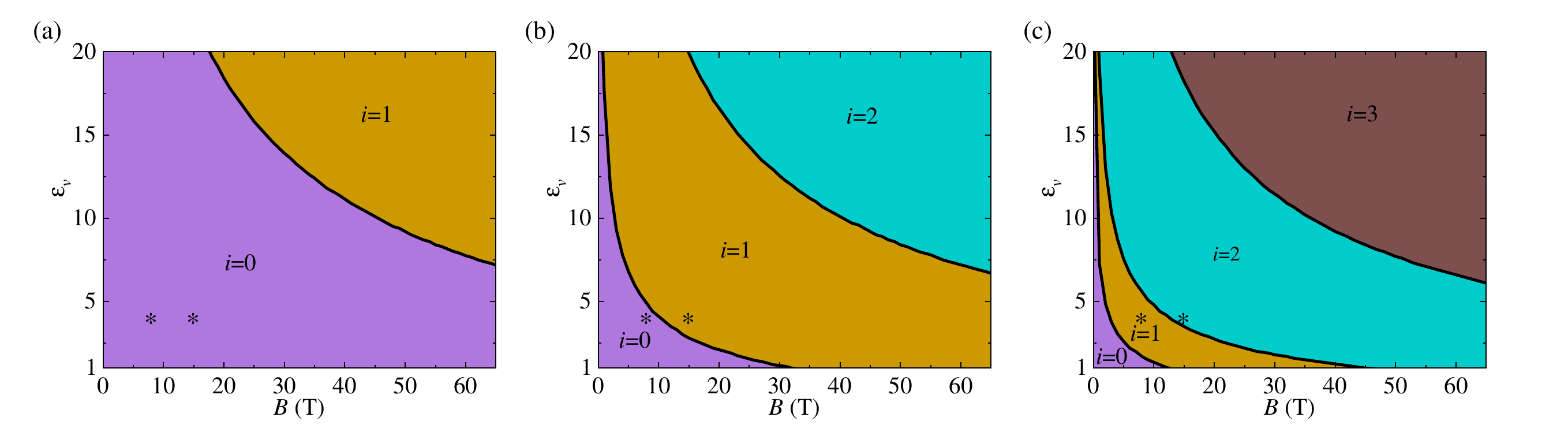}
	\caption{(Color online) The phase diagrams of the dominant electron-hole pair $\{n_e=n_h=i\}$ in the parameter space spanned by the magnetic field $B$ and the relative dielectric constant $\epsilon_v$.  (a)-(c) show the results of the $2s$, $3s$ and $4s$ states, respectively.  The different dominant components $i$ are labeled in different colors.  The two asterisks marked in each figure correspond to the parameters chosen in Fig.~\ref{Fig4}(a).}    
	\label{Fig5} 
\end{figure*}

In the Landau quantization space, we can analyze the exciton state composition.  That is, from the wave function $\Psi_{nl}$, the weight $|\phi_{nl}^i|^2$ of the free electron-hole pair $\{n_e=i+l, n_h=i\}$ can be extracted~\cite{D.V.Tuan2025}.  The composition of different exciton states is  displayed in Fig.~\ref{Fig4} as a function of the hole LL index $i$.  Since the exciton states are mainly composed of the low-energy pairs, the weights are plotted from $i=0$ to $i=15$.  

Intuitively, without the Coulomb interactions, the exciton state $nl$ includes only the free electron-hole pair $\{n_e=n+l-1,n_h=n-1\}$, as the non-interacting exciton wavefunction is just a product of the electron and hole LL wavefunctions; the Coulomb interactions can mix more electron-hole pairs into the exciton state.  
In Fig.~\ref{Fig4}(a), when the magnetic field $B=8$ and $15$ T, the largest component of both the $1s$ and $2s$ states lies at $i=0$, indicating that the $\{n_e=n_h=0\}$ pair dominates the two states. 
This means that the Coulomb interactions do not change the dominant component of the $1s$ state, but will shift that of the $2s$ state from $i=1$ to $0$.  

In Fig.~\ref{Fig4}(a) insets, the weights $|\phi_{ns}^i|^2$ for the $ns$ states are shown with the index $i$ up to the cutoff $i=300$.  
In Fig.~\ref{Fig4}(a) upper inset, when $B=15$ T, the weight $|\phi_{1s}^{300}|^2$ of the ground state $1s$ at $i=300$ reaches $10^{-6}$; while the weights $|\phi_{2-4,s}^{300}|^2$ of the higher exciton states are vanishing.  As the nonvanishing $|\phi_{1s}^i|^2$ can extend to much larger $i$, the ground state includes more electron-hole pairs than other states and thus behaves more ``extensive" in the Landau quantization space.  
By comparison, in Fig.~\ref{Fig4}(a) lower inset, when $B=8$ T, the weight $|\phi_{1s}^{300}|^2$ increases to $10^{-5}$, and the weights $|\phi_{2-4,s}^i|^2$ of other exciton states vanish at higher $i$.  
This means when the magnetic field decreases, more electron-hole pairs will be included in the exciton states, which can help understand why in the Landau quantization space, the convergence of the lower levels is poor under weak magnetic fields.  

Moreover, in Fig.~\ref{Fig4}(a) of the $3s$ $(4s)$ state, the dominant component is $i=0$ $(1)$ when $B=8$ T, and will shift to $i=1$ $(2)$ when the magnetic field increases to $B=15$ T.  
Similar dominant component shift can also be found in other exciton states.  In Fig.~\ref{Fig4}(b) of the $4p$ state, the dominant component is $i=0$ when $B=5$ T, and will shift to $i=1$ when $B=10$ T; in  Fig.~\ref{Fig4}(c) for the $4d$ state, the dominant component is $i=0$ when $B=25$ T, and will shift to $i=1$ when $B=35$ T.  These results indicate that with increasing $B$, the dominant component shift is ubiquitous in the exciton states.  

To further explore the physics underlying the exciton state composition, we take the $s$ states as an example and investigate how their dominant electron-hole pair components $\{n_e=n_h=i\}$ are affected by the magnetic field and Coulomb interactions.  Here, we use the relative dielectric constant $\epsilon_v$
to control the interaction strength~\cite{I.Kylanpaa}.  Experimentally, the dielectric environment can be effectively modulated by encapsulating the 2D flakes with different materials~\cite{A.V.Stier2016a}.  
In the parametric space $(B,\epsilon_v)$, we calculate the phase diagram of the dominant electron-hole pair.  The results of the $2s$, $3s$ and $4s$ states are displayed in Figs.~\ref{Fig5}(a)-\ref{Fig5}(c), respectively.  Note that $\epsilon_v$ begins from its lower limit $\epsilon_v=1$, which corresponds to the vacuum case.  In the $ns$ state, as the dominant component can shift from the lowest pair of $i=0$ up to the highest pair of $i=n-1$, the total $n$ different dominant components are included in each phase diagram.  More observations are given as follows. 

(i) In the limit $\epsilon_v\rightarrow\infty$, the interaction strength tends to be zero.  Although such a limiting case is not seen in the phase diagram, the asymptotic behavior that the component $i=n-1$ dominates the $ns$ state can still be expected.  This is consistent with the above intuitive analysis of the case without the Coulomb interactions.   
 
(ii) With decreasing $\epsilon_v$, the interaction strength increases.  
At a fixed $B$, the increasing interaction will drive the dominant component to shift from $i=n-1$ to $i=0$ successively.
This is because the low-$i$ electron-hole pairs can have a more compact wavefunction and allow the attractive Coulomb interactions gain more energy.  For the strong interactions, the gain in the Coulomb interaction energy outweighs the kinetic energy penalty of moving away from the non-interacting state, which makes the lower $i$ component dominant.  
When $B$ grows, the critical $\epsilon_v$ separating the different components gradually decreases.
Note that the two asterisks marked in each figure correspond to the two sets of parameters $(B,\epsilon_v)$ chosen in Fig.~\ref{Fig4}(a), from which the dominant component shift in the $3s$ and $4s$ states can be understood.  

(iii) When $\epsilon_v=1$, the interaction strength reaches its maximum.  For the field range $B\sim(0, 65\text{ T})$, in Fig.~\ref{Fig5}(a) of the $2s$ state, the dominant component remains as $i=0$.  By comparison, in Fig.~\ref{Fig5}(b) of the $3s$ state, the dominant component will shift from $i=0$ to $1$ when $B>32$ T; in Fig.~\ref{Fig5}(c) of the $4s$ state, the dominant component will shift to $i=1$ when $B>13$ T and further to $i=2$ when $B>46$ T.  

Based on these observations, we reveal that in the $ns$ state, the magnetic field will drive the dominant component to the electron-hole pair of $i=n-1$, while the Coulomb interactions will drive the dominant component to the pair of the lower $i$ that owns the lower energy; their competitions determine the dominant component behaviors in Fig.~\ref{Fig5}.  
We suggest that the roles played by the magnetic field and Coulomb interactions in determining the dominant component can also be found in other exciton states, $p$, $d$, $f$, $\cdots$.

\subsection{Exciton stability under magnetic fields}

\begin{figure}
	\includegraphics[width=8cm]{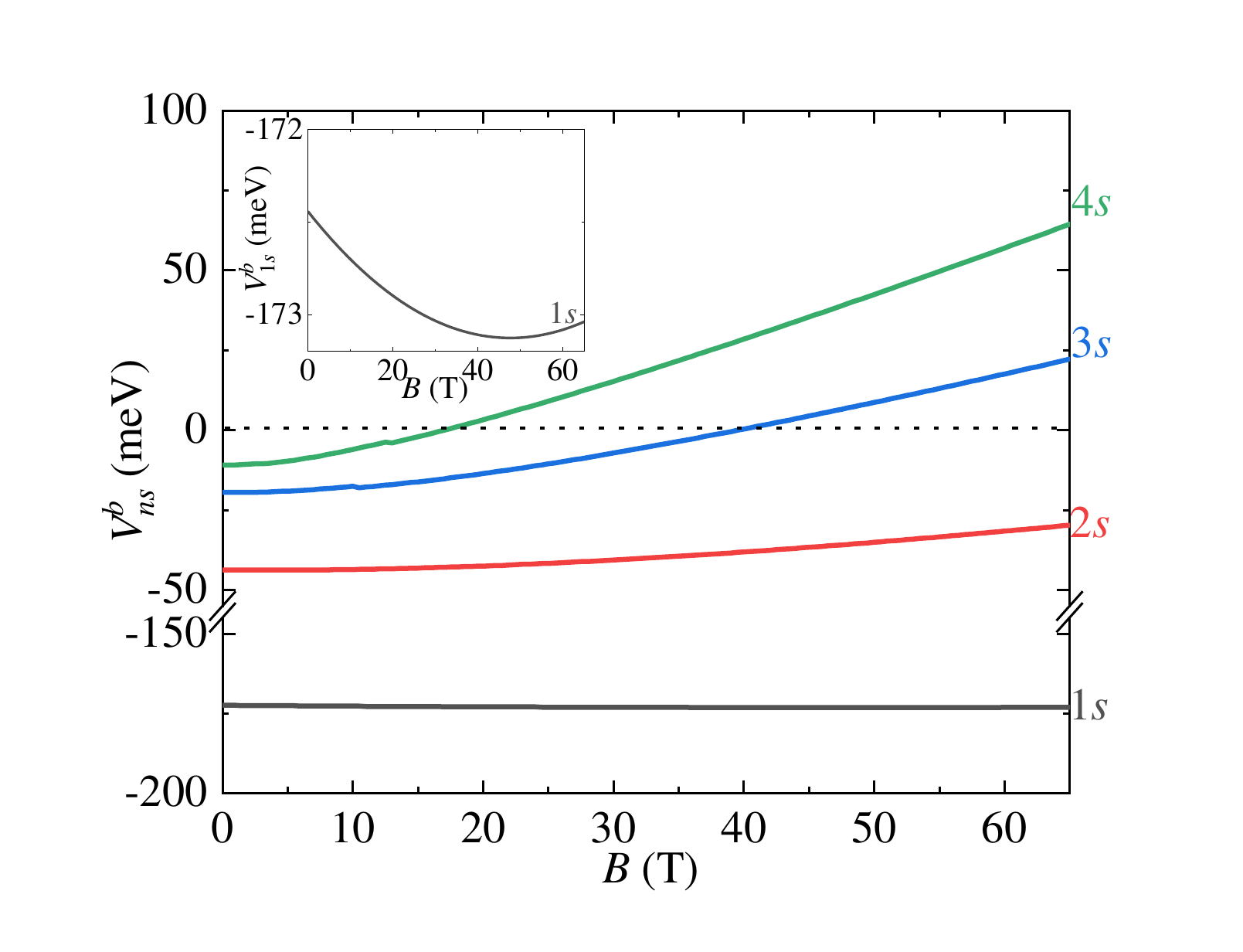}
	\caption{(Color online) The Coulomb binding energy $V_{ns}^b$ vs the magnetic field $B$ for the $ns$ states.  Here the binding energy $\varepsilon_{ns}$ is obtained from the real space calculations.  The inset shows the enlarged plot for the $1s$ state. }    
	\label{Fig6} 
\end{figure}

Although the exciton is charge neutral and is not subjected to the Lorentz force, the relative motion of the electron and hole is still affected by the magnetic field, as seen in Eq.~(\ref{Seq}).  

To explore the exciton stability under magnetic fields, we define the Coulomb binding energy $V_{nl}^b$ as the exciton energy $\varepsilon_{nl}$ subtracting the dominant electron-hole energy $\varepsilon_{nli}^d$, with $\varepsilon_{nli}^d=(i+l+\frac{1}{2})\hbar\omega_e+(i+\frac{1}{2})\hbar\omega_h$, 
\begin{align}
V_{nl}^b=\varepsilon_{nl}-\varepsilon_{nli}^d.  
\end{align}
The dependence of the exciton energy on the magnetic field can be written as   $\varepsilon_{nl}(B)=\varepsilon_{nl}(0)+\Delta\varepsilon_{nl}^{\text{dia}}(B)+\varepsilon_{nl}^{\text{Z}}(B)$, in which $\varepsilon_{nl}(0)$ denotes the zero-field energy, $\Delta\varepsilon_{nl}^{\text{dia}}(B)$ is the diamagnetic shift, and $\varepsilon_{nl}^{\text{Z}}(B)=l\mu_B^{\text{ex}} B$ is the Zeeman shift. 
We adopt a simple criterion to judge the exciton stability: when $V_{nl}^b<0$, the exciton state $nl$ is stable; contrariwise, when $V_{nl}^b>0$, the exciton state is unstable.  In Fig.~\ref{Fig6}, we plot $V_{ns}^b$ as a function of $B$ for the $ns$ states, where the dominant component index index $i$ of $\varepsilon_{nli}^d$ is obtained from Fig.~\ref{Fig5} with $\epsilon_v=3.97$.  Note that in the $1s$ and $2s$ state, the dominant components are $i=0$ and remain unchanged to the magnetic field.  

Consider the ground state $1s$.  Under weak magnetic fields, we have 
$\Delta\varepsilon_{1s}^{\text{dia}}=\sigma_{1s}B^2$ and $\sigma_{1s}=0.31$ $\mu$eV/T$^2$, $\varepsilon_{1s}^{\text{Z}}=0$, and $\varepsilon_m=0.02895 B$ meV.  As the lowest cyclotron energy is much larger than the diamagnetic shift,  $\varepsilon_m\gg\Delta\varepsilon_{1s}^{\text{dia}}$, the Coulomb binding energy decreases with the magnetic field, meaning that the exciton binding will get strengthened.  Under strong magnetic fields, the diamagnetic shift becomes linear with $B$,  $\Delta\varepsilon_{1s}^{\text{dia}}=k_{1s}B+b_{1s}$ and $k_{1s}=0.032$ meV/T. 
Now the diamagnetic shift grows more quickly with $B$ than $\varepsilon_m$ and the exciton becomes loosely bound.  But as the Coulomb binding energy $V_{1s}^b$ is still negative, the ground state $1s$ remains stable.  This is clearly seen in Fig.~\ref{Fig6} inset.  Similar analysis can be made to other exciton states.  
In Fig.~\ref{Fig6}, we observe that for the $3s$ state, $V_{3s}^b>0$ when $B>40$ T; for the $4s$ state, $V_{4s}^b>0$ when $B>17$ T, indicating that these exciton states will become unstable when the magnetic field crosses a critical value.  These results can explain the experimental observations that the $1s$ and $2s$ exciton states are stable and can persist to the strong magnetic field, while the $3s$ and $4s$ states become blurred and vanish under strong magnetic fields~\cite{A.V.Stier2018, E.Liu2019}.

\section{Discussions and Summaries}

To summarize, in this paper, we have made a comparative study of the 2D bound excitons in monolayer WSe$_2$ under a magnetic field in both real space and Landau quantization space.  
While real space focuses on the relative motion of the electron and hole, the Landau quantization space is based on respective cyclotron motion of the electron and hole in a magnetic field, meaning that  the two spaces have different physical pictures. 
In calculating the exciton energy levels, the real space method has the advantage of being directly applicable down to arbitrarily weak magnetic fields, while the Landau quantization method requires an increasing larger basis when the magnetic field decreases, but can help analyze the exciton composition. 
The two spaces complement each other and can provide supportive perspectives in understanding the exciton behaviors. 

We obtain the exciton energy spectrum and diamagnetic shift in the two spaces, with the results agreeing well with each other.   
Compared to Ref.~\cite{D.V.Tuan2025}, in which the exciton energy levels are only solved in the Landau quantization space, our work gives a cross-validation of the two methods.  
More importantly, according to the exciton state composition analysis, we reveal the roles played by the magnetic field and Coulomb interactions in determining the dominant component, which provide useful perspectives on how the external factors will modulate the exciton structures. 

In the dipole approximation, the optical matrix element for creating the exciton state $nl$ from the vacuum (with all valence bands occupied and all conduction bands unoccupied) is ${\cal M}_{nl}=\langle nl|H'|0\rangle$.  Here the interaction Hamiltonian $H'=-{\boldsymbol P}\cdot{\boldsymbol E}$, in which ${\boldsymbol P}$ is the polarization vector and $\boldsymbol E$ is the electric field strength.  
In the Landau quantization space, we have $|nl\rangle=\sum_i\phi_{nl}^i|i+l,i\rangle$, in which $\phi_{nl}^i$ denotes the wavefunction component and $|i+l,i\rangle$ is the state of the electron-hole pair $\{n_e=i+l,n_h=i\}$.  Then we have ${\cal M}_{nl}=\sum_i(\phi_{nl}^i)^*{\cal M}_{i+l,i}$, with ${\cal M}_{i+l,i}=\langle i+l,i|H'|0\rangle$ denoting the optical matrix element for the free electron-hole pair.  
In the optical absorption and magneto-optical response experiments, the peak intensity (or called the oscillator strength) is related to the optical matrix element as $f_{nl}\propto|{\cal M}_{nl}|^2=|\sum_i (\phi_{nl}^i)^*{\cal M}_{i+l,i}|^2$, which is given as a coherent summation of different components.  Thus, when crossing the transition, the dominant component shift will be accompanied by the variation of the peak intensity.   

Besides WSe$_2$, we believe that the Landau quantization method is also valid in analyzing the exciton state composition in other monolayer TMDs~\cite{A.V.Stier2016b, T.C.Berkelbach, A.Chernikov, I.Kylanpaa}, MoS$_2$, MoSe$_2$, and WS$_2$, as well as in TMD heterojunction moir\'e superlattices~\cite{D.Chen, R.Xiong, C.Jin}.  When the massive Dirac model is used to describe the excitons in monolayer TMDs~\cite{M.Donck2018, M.Donck2019}, in which the spin-orbit coupling and the valley degree of freedom are included, the Landau quantization method can also be applied and needs more investigations. 

\begin{figure}
	\includegraphics[width=9.2cm]{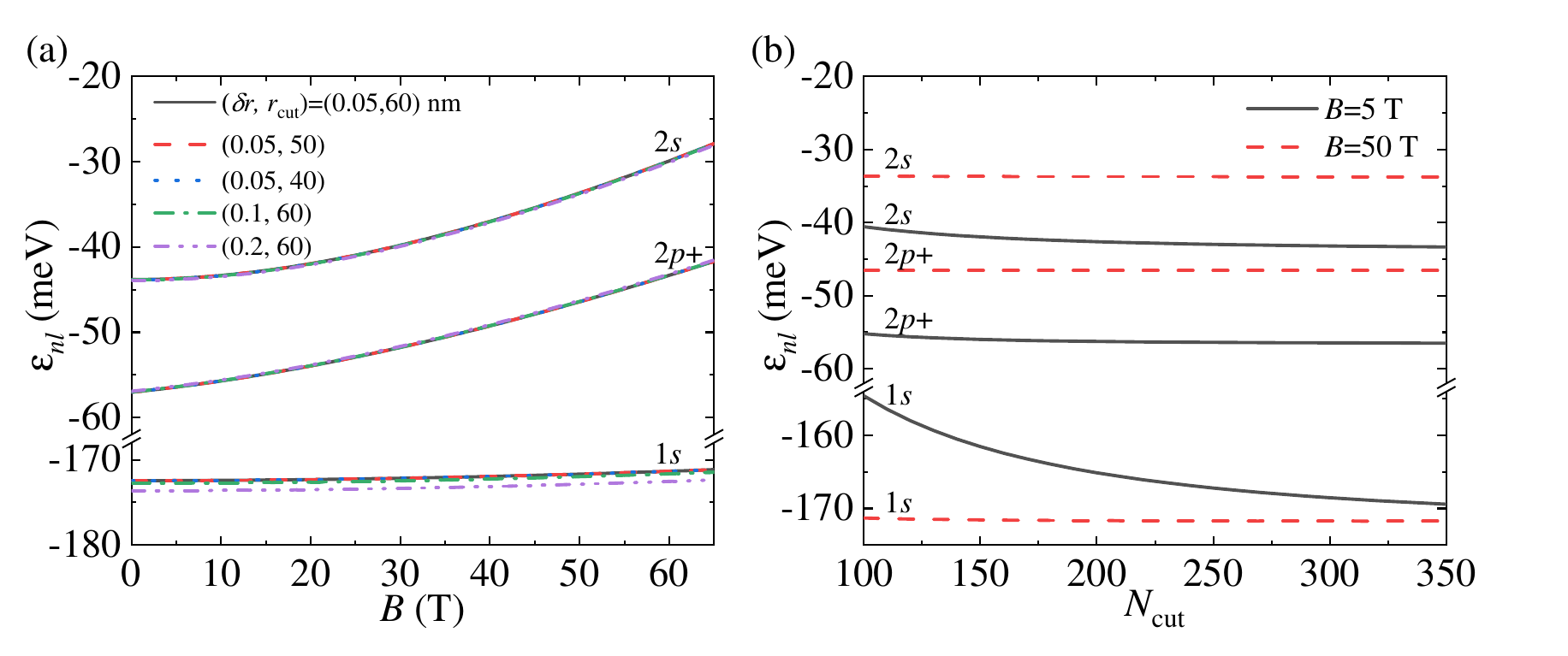}
	\renewcommand{\thefigure}{A1}	
	\caption{(Color online) The convergence of the three lowest levels of the bound excitons.  (a) The energy levels obtained from the real space calculations vs the magnetic field $B$ for a set of the grid separation and the cutoff grid size $(\delta r, r_{\text{cut}})$.  (b) The energy levels obtained from Landau quantization space calculations vs the matrix dimension cutoff $N_{\text{cut}}$ when $B=5$ T and $50$ T.}  
	\label{FigA1}
\end{figure}

\section{Acknowledgments} 

This work was supported by the Natural Science Foundation of China (Grant No. 11804122).

\section{Data availability} 

The data that support the findings of this paper are openly available in the GitHub repository at Ref.~\cite{data}. 
The repository includes all relevant datasets and Fortran codes.

\section{Appendix}

Here we provide more details for the convergence of the numerical calculations.   

Figure~\ref{FigA1}(a) shows the three lowest levels of the bound excitons obtained in real space, which are given as a function of the magnetic field $B$ for a set of grid separations and the cutoff grid size $(\delta r, r_{\text{cut}})$.  
We see that $\varepsilon_{1s}$ converges quickly with smaller $\delta r$ and larger $r_{\text{cut}}$; while $\varepsilon_{2s}$ and $\varepsilon_{2p+}$ show negligible variation for different $(\delta r, r_{\text{cut}})$.  These results indicate that in real space, the convergence of the energy levels can be quickly reached.  To make the results more reliable, we choose $(\delta r,r_{\text{cut}})=(0.05,60)$ nm in the main text.  

Figure~\ref{FigA1}(b) shows the three lowest levels obtained in the Landau quantization space, which are given as a function of the matrix dimension cutoff $N_{\text{cut}}$ when $B=5$ T and $50$ T.  
At the weak magnetic field $B=5$ T, we see that $\varepsilon_{1s}$ decreases slowly with $N_{\text{cut}}$.  To get an accurate result of $\varepsilon_{1s}$, one needs to set $N_{\text{cut}}$ to be large enough.  For $\varepsilon_{2s}$ and $\varepsilon_{2p+}$, the convergence is reached at a smaller $N_{\text{cut}}$.  
By comparison, at the strong magnetic field $B=50$ T, we observe that the energy levels are almost unchanged with $N_{\text{cut}}$.  
We conclude that in the Landau quantization space, the convergence is poor for the lower energy levels and weak magnetic field, but can be quickly reached for the higher energy levels and the strong magnetic field.  Considering our computational resources, we set $N_{\text{cut}}=300$ in the main text.

\end{document}